# Soundiation: A multi-functional GUI-based software in evaluation of the acoustophoresis by the acoustic radiation force and torque on arbitrary axisymmetric objects

Tianquan Tang[a,b,*], Lixi Huang[a,b]

[a]*Department of Mechanical Engineering, The University of Hong Kong, Pokfulam, Hong Kong SAR, China*
[b]*Lab for Aerodynamics and Acoustics, HKU Zhejiang Institute of Research and Innovation, 1623 Dayuan Road, Lin An District, Hangzhou, China*

Corresponding email: tianquan@connect.hku.hk

Acoustic radiation force and torque arising from wave scattering are commonly used to manipulate micro-objects without contact. We applied the partial wave expansion series and the conformal transformation approach to estimate the radiation force and torque exerted on an axisymmetric particle. Meanwhile, translational and rotational transformations are required to keep the coordinate system consistent [1]. Although these theoretical derivations have been well established, coding the required systems, including generation of the wave function, implementation of the transformations, calculations between modules, etc., is non-trivial and time-consuming. Here, a new open-source, MATLAB-based software, called Soundiation, is provided to address the radiation force and torque while supporting the dynamic prediction of non-spherical particles. The implementation is basically generic, and its applicability is demonstrated through the validation of numerical methods. Furthermore, a graphical user interface is provided so that it can be used and extended easily.

# 1. Motivation and significance

Acoustic tweezers use acoustic radiation force and torque to manipulate matter without contamination. The radiation force and torque are exerted on objects because of the momentum transfer that results from acoustic scattering effects of the wave-particle interaction [2][3][4]. Since they are able to perform biocompatible, contact-free, and precise translation and rotation of micro-objects, this software greatly facilitates studies in engineering and biotechnology, including holographic acoustic tweezer [5][6], microorganism morphological interrogation [7], and microsurgery [8].

For Rayleigh objects, where the scattering effect is negligible, the radiation force and torque on the particles are obtainable according to Gorkov's theory [9]. Beyond the Rayleigh regime, the radiation force and torque on the spherical particles can be solved with the help of partial wave expansion series [10][11]. However, the above methods are assumed the manipulated particles are small or spherical. In reality, geometric asymmetry naturally exists in objects, such as *erythrocyte* [12] and *C. elegans* [13]. An alternative to evaluate the radiation force and torque on the non-spherical particles is the use of numerical methods [12][14], while it is limited by high computational cost and time-consuming.

To expand the range of applications while improving the efficiency in computations, in our previous studies, we established a general analytical framework to estimate radiation force and torque on any axisymmetric particles induced by traveling plane waves and user-customized transducer arrays [15]. Here, in this contribution, we present a high-performance and user-friendly graphical user interface (GUI), Soundiation, for the radiation force and torque, thereby the acoustophoretic process. It is worth mentioning that the major theories applied in this software include partial wave expansion method [16], translation addition theorem [17], and conformal mapping approach [18]. The software is general in terms of particle geometries, particle sizes, and frequency range. Additionally, the user is able to decide the arrangement of the transducer configurations. The host fluid and particle boundary conditions also remain flexible. Some potential functionalities of this software are concluded as: (1) designing the non-spherical particle and generating its three-dimensional geometric data, which can be conveniently imported to other commercial numerical software (e.g., COMSOL Multiphysics) for further research; (2) visualizing the specified plane wavefield or piston-like wavefield from a user-customized transducer array; (3) estimating the acoustic radiation force and torque exerted on the particle designed in (1), and the incident wavefield is specified in (2); (4) predicting and visualizing the time-variant dynamic process of the particle designed in (1) under the incident wavefield given in (2), i.e., illustrating the time-variant acoustophoretic process of the particle.

A brief overview of the software, Soundiation, is described in section 2 (more explanations in how to use the software can refer to the user manual given in the "Soundiation-Acoustophoresis ./docs/user_manual" folder). In section 3, a series of radiation force and torque under different scenarios have been estimated and validated by the three-dimensional numerical simulations (the numerical model used to validate is given in the "Soundiation-Acoustophoresis ./docs/COMSOL") folder), making the results acceptable in other applications. Additionally, we demonstrate an example, the acoustophoretic process of an ellipsoid particle above a transducer array, to illustrate the ability of the software in prediction of dynamic processes. Finally, the impact and conclusions are described in sections 4 and 5, respectively.

## 2. Software description

### 2.1. Software principle

The reader is recommended to refer to our previous studies for more theoretical details in the determinization of mapping coefficients $c_n$ for different geometries [1], the coordinate transformation [1], the translation addition theorem in the transducer array system [15], and the radiation force and torque for the acoustophoretic process of particles [1].

### 2.2. Software architecture

Here, a brief summary of the major features is presented below. For user-friendly considerations, the software is demonstrated as a GUI, which has been tested with MATLAB 2010a and above versions. When we run the GUI ("main_interface.m"), a set of default parameters is given, while the users can change all the parameters based on their needs. The source codes can be found in the "Soundiation-Acoustophoresis ./src" folder, while the explanation of parameters required in the graphical user interface (GUI) can be found in the "Soundiation-Acoustophoresis ./docs/user_manual" folder. As shown in Fig. 1, the GUI of Soundiation is divided into three main panels, "Visualization", "Calculation controls", and "Functions".

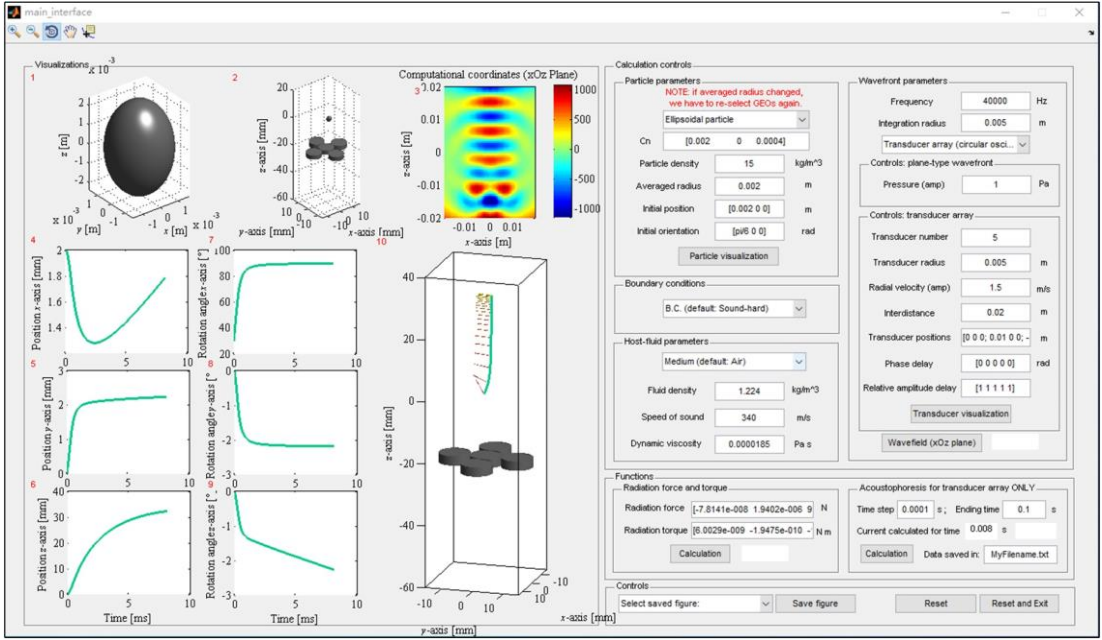

**Figure 1:** Graphical user interface (GUI) of Soundiation.

The first penal, "Visualization", is designed to visualize the optional results, including geometry of particle, transducer-particle system, incident wavefield, and time-variant trajectory of particle driven by the acoustic incident wavefields (i.e., acoustophoretic process). The "Visualization" panel has ten regions, labeled by 1 to 10. Note that all figures illustrated in the "Visualization" panel can be saved using the "Save figure" button.

The second panel, "Calculation controls", defines the computational conditions, geometric information, and material properties. Specifically, there are four kinds of controlling parameters. Firstly, the geometry and physical properties of a particle must be defined. Secondly, the boundary condition to the particle surface should be selected, including the sound-hard (Neumann) and the sound-soft (Dirichlet) boundary conditions. Thirdly, the medium that immersed the particle should be selected. Finally, the incident wavefield should be specified by giving the complex pressure amplitude for the plane wavefront or the transducer parameters for the wavefront of the transducer array.

The final panel, "Functions", provides two optional calculations based on the parameters set in the "Calculation controls" panel. The first sub-panel provides functionality to directly predict the acoustic radiation force and torque exerting on the user-specified particle under user-customized incident wavefront. The second sub-panel is designed to compute a dynamic process for the scenario that any axisymmetric, irregular particle levitates above a user-customized transducer array. The time-variant trajectories, including translational and rotational information, are visualized in the "Visualization" panel.

## 2.3. Software functionalities

Soundiation is designed to efficiently estimate the radiation force and torque on various particles under a sound wavefield. Combined with the viscous drag force and torque [19][20], the spatio-temporal acoustophoresis of these particles can be predicted. It should be emphasized that the shape of particles, including spherical or arbitrary axisymmetric geometries, can be designed by the user. The incident wavefield also supports user-customized. These geometric degrees of freedom greatly improve the application of this software in a wider range of scenarios. More specifically, the functionalities and instructions are introduced in detail below.

### 2.3.1. Particle-transducer system and incident wavefield visualization

A three-dimensional graph of the manipulated non-spherical particle can be designed by specifying the mapping coefficients $c_n$; the corresponding geometry is illustrated in the "Visualization" panel (labeled by 1, Fig. 1). The surface coordinates are automatically saved in the "particle_data.stl" file, which facilitates the import of particle geometry into other numerical software (e.g., COMSOL Multiphysics) and generates three-dimensional structures to perform numerical calculations if needed. The geometric differences are captured by different combinations of the mapping coefficients $c_n$, while the general geometric size can be stretched by changing the parameter of "Averaged radius". It should be emphasized that the first coefficient of the mapping coefficients is equivalent to the averaged radius. After clicking the "Particle visualization" button, the particle geometry is plotted in the "Visualization" panel.

Additionally, for the case of using transducer arrays, three-dimensional visualization of the particle-transducer system is available. Note that to graph the result, the user needs first to initialize and parameterize the particle and the transducer array. Particle "Initial position" represents the values that the particle deviates from the coordinate origin (0,0,0), while particle "Initial orientation" gives the angles that the particle rotates along the $x$-, $y$-, and $z$-axes, respectively. For example, if "Initial position = [0.002, 0, 0] mm" and "Initial orientation = [$\pi/6$, 0, 0] rad", the particle should be placed at $x = 2$ mm, $y = 0, z = 0$ and be rotated along $x$-axis for $\pi/6$ rad. Similarly, all transducers should be parameterized. The number, radius, and positions of transducers should be specified in "Transducer radius", "Transducer number", and "Transducer positions". A $N \times 3$ matrix is required to define the position, where $N$ corresponds to the number of transducers, and their central positions are recorded in the row vectors. The first transducer is defined as the probe transducer, whose Cartesian coordinates are set to (0,0,0), while the coordinates of the other transducers are given relative to the probe transducer. As a result, any desirable distribution of transducers is possible. Note that the "Transducer positions" given above are under the premise that the probe transducer coincides with the coordinate origin. To change the vertical position of the transducer array, the parameter "Interdistance" is introduced, which indicates a value that translates the array along $-z$-axis. After clicking the "Transducer visualization" button, the particle-transducer system is depicted in the "Visualization" panel (labeled by 2, Fig. 1).

Soundiation uses the partial wave expansion series to describe incident wave propagation in the invisous medium [16]. The method gives either a plane wavefield or a summation wavefield from multiple circular radiators (or transducers). It should be emphasized that the vibration of each circular radiator can be regarded as a piston-like vibrator whose wavefield is well-known [21], and thus the summation wavefield can be derived with the help of the translation addition theorem [1][15]. Before visualizing the incident wavefield, the user needs to first parameterize the wavefront parameters in the GUI. Apart from the "Frequency", the user is required to define the pressure amplitude in "Pressure (amp)" for the plane traveling wave, while the amplitude of radial velocity and the array delay (phase and relative amplitude delays) in "Radial velocity (amp)", "Phase delay", and "Relative amplitude delay". In this way, once the incident wavefront parameters have been specified, a pressure wavefield on $xOz$ plane is available to give an impression of the incident wave patterns in "Visualization" panel (labeled by 3, Fig. 1) by clicking the "Wavefield ($xOz$ plane)" button.

### 2.3.2. Acoustic radiation force and torque on non-spherical particles

This module calculates radiation force and torque exerted on the axisymmetric particles excited by a plane wave or a user-customized transducer array.

Parameter definitions of particles and incident wavefield are explained in section 2.3.1. In order to estimate the radiation force and torque, except to parameterize the particles and the incident wavefront, the user needs to specify the boundary condition (around the particle surface) and the medium properties. The medium properties can be simply defined by selecting an in-house medium from the popup menu in the "Host-fluid parameters" sub-panel. Physically, the sound-soft (Dirichlet) and the sound-hard (Neumann) boundary conditions require that the total potential vanishes and the normal particle velocity vanishes, respectively, on the scatterer surface. A detailed derivation can be found in our previous work [1]. Nevertheless, from the user point of view, there is no need to enter into the detail of all the related formulas. The GUI provides a convenient way to build the boundary condition around the scatterer surface by selecting a desirable one from the popup menu in the "Boundary conditions" sub-panel.

After completing all the above settings, the corresponding radiation force and torque will be presented in the edit boxes "Radiation force" and "Radiation torque" by clicking the "Calculation" button in the "Radiation force and torque" sub-panel (the computation will take a few seconds). The data is given in the format of a $1 \times 3$ row vector, representing the radiation force and torque components in the $x$-, $y$-, and $z$-directions, respectively.

### 2.3.3. Dynamic process: particle acoustophoresis

The software is able to predict the spatio-temporal trajectories of a user-defined particle under the acoustic field, i.e., the acoustophoretic process. Consider that the particle is driven by acoustic radiation force and torque, while delayed by the drag force and torque due to the viscous stresses and shear stresses on the particle surface [19][20]. Since the drag force and torque is proportional to the particle relative velocity and angular velocity, which source from the radiation force and torque. Therefore, the time-dependent process is obtainable. Once the initial status of the particle is given, the radiation force and torque can be evaluated. As a result, the particle velocity and angular velocity (i.e., the drag force and torque) can be solved following the equation of motions. Then, these velocities are applied to renew the particle position and orientation for the next time step [1].

To carry on the prediction of the acoustophoretic process, except to the settings given in sections 2.3.1 and 2.3.2 for the radiation force and torque for the initial status, the user needs to specify further an ending moment ("Ending time") to stop the prediction and a time step ("Time step") to continue the computation. Note that we end the predictions when the changes of the positions and the rotation angles among two adjacent time steps are less than 5%. It should also be mentioned that the in-house wavefield radiated from a circular oscillator is based on the far-field assumption [21]. Therefore, we stop the calculations once the vertical distance (or interdistance) between the particle and the transducer array is smaller than 10 mm.

Prediction begins after clicking the "Calculation" button in the "Acoustophoresis for transducer array" sub-panel. And the predicted translational and rotational trajectories are depicted in the "Visualization" panel. The time-variant translational trajectories along $x$-, $y$-, and $z$-axes are plotted in the "Visualization" panel, labeled by 4, 5, and 6 (Fig. 1), respectively. The time-variant rotational tendencies along $x$-, $y$-, and $z$-axes are graphed in blank regions 7, 8, and 9 (Fig. 1), respectively. Correspondingly, a three-dimensional result is given in region 10 (Fig. 1), where the solid line indicates the translational trajectory while the arrows demonstrate the particle orientation (axisymmetric axis). The color of the arrow is used to represent the increase of time (red-yellow color spectrum). The time intervals represented by any adjacent arrows are the same. We apply a total of 20 arrows to continuously demonstrate the rotational property of the particle from the start of the calculation (red arrow) to the end of the calculation (yellow arrow). Additionally, the control parameters and the corresponding dynamic process (i.e., the translational positions and the rotational angles at different time steps) are recorded in a user-defined ".txt" file automatically. By default, the file name is "Myfilename.txt".

## 3. Illustrative Examples

Here, we predict the radiation force and torque on different particles using the Soundiation software (GUI), followed by comprehensive validations with numerical results from COMSOL. Afterward, we demonstrate an example to predict the spatio-temporal acoustophoretic process of a non-spherical particle above a transducer array.

### 3.1. Validations: radiation force and torque on various particles

The particles can be created by specifying proper mapping coefficients and clicking the "Particle visualization" in the "Particle parameters" sub-panel, as shown in Fig. 1. The corresponding particle geometry will be present in region 1 of the "Visualization" panel; meanwhile, the geometric data is automatically saved in the "particle_data.stl" file. The data file can be directly imported into COMSOL to facilitate the establishment of the numerical model, and if required, to validate the predictions.

For the first example, consider that a time-harmonic acoustic plane wave interacts with different axisymmetric particles (including cone and cylindrical particles with mapping coefficients of $c_n$ = [0.002 0 0 0.00025] and $c_n$ = [0.002 0 -0.0005 0 -0.00025], respectively) at different particle orientations. The designed cone and cylindrical particles are illustrated in Figs. 2(a) and 3(a). We assume that the particles are all rigid, corresponding to the sound-hard or Neumann boundary condition, and their averaged radius is set to 2 mm. The plane wave is propagating in the air, operating at a frequency of 40 kHz and a pressure amplitude of 1 Pa. To demonstrate and validate the software property in predicting the radiation force and torque under a plane wavefield, we carry on a series of comparisons between the predicted results and the numerical simulations as shown in Figs. 2 and 3. In all cases, the particle position is set to [0 0 0]. Differently, the particle orientation varies along $x$-axis, that is $\theta_{\text{rot},x} \in [0, 90°]$.

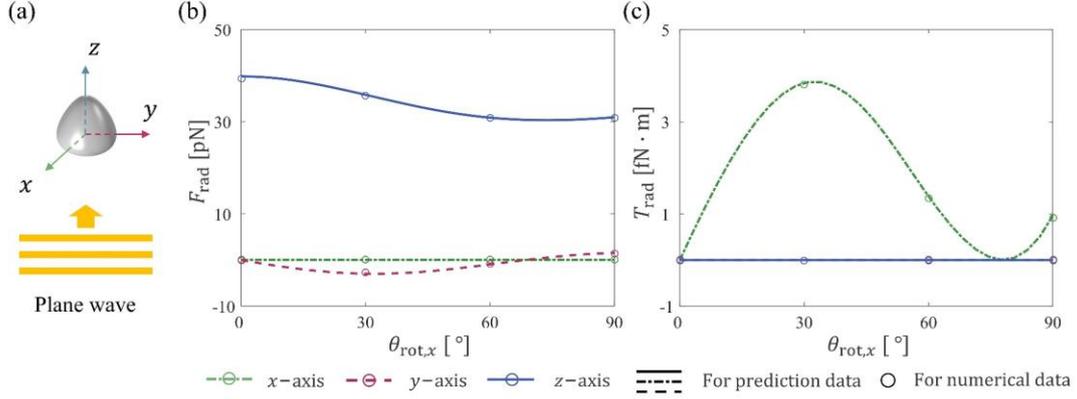

**Figure 2:** Theoretical and numerical calculations of the acoustic radiation force, $\vec{F}_{\text{rad}}$, and torque, $\vec{T}_{\text{rad}}$, resulting from an incident plane wave (amplitude of 1 Pa) acting on a sound-hard cone particle (averaged radius of 2 mm) as a function of the particle orientation $\theta_{\text{rot},x} \in [0, 90°]$. (a) Schematic diagram of a plane wave interacting with the cone particle under the Cartesian coordinate system. (b) and (c) Comparison plots for the radiation force and torque, respectively. The green dashed-dot lines, red dashed lines, and the blue solid lines indicate the prediction data along $x$-, $y$-, and $z$-axes, while the circle marks represent the corresponding results based on the full three-dimensional numerical simulations.

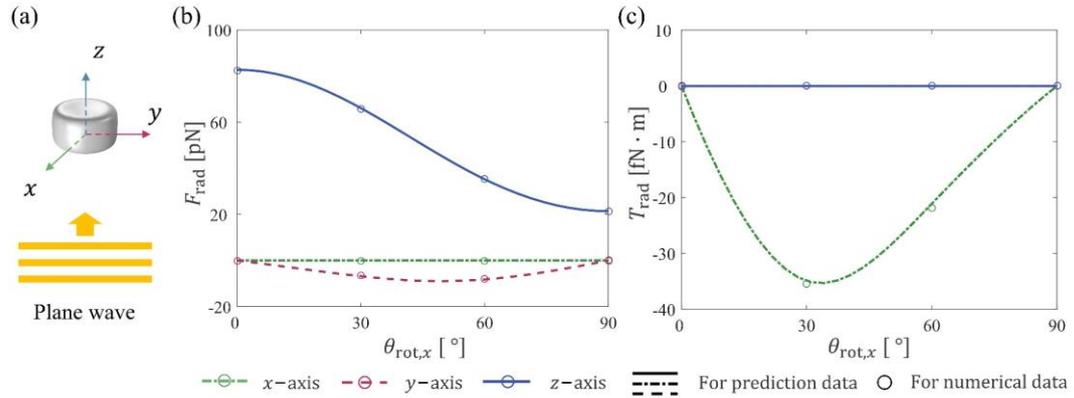

**Figure 3:** The same as in Fig. 2, but the scatterer is changed to a cylindrical particle.

For another example, we consider that a diamond particle ($c_n = [0.002 \ \ 0 \ \ 0 \ \ 0 \ \ 0.0002]$) is levitated above a transducer array. The particle is rigid (sound-hard or Neumann boundary condition), and its averaged radius is also set to 2 mm. The diamond particle used here is shown in Fig. 4(a) (or Fig. 5(a)). The sound waves radiated from the circular radiators, the amplitude of the radial velocity of 1.5 m/s, are propagating in the air, operating at a frequency of 40 kHz. The transducer array consists of four transducers with radii of 5 mm. The transducer configuration, phase delay, and relative amplitude delay are arranged as shown in Fig. 4(a) or 5(a). Specifically, we define "Transducer positions" as $[0\ 0\ 0;\ 0.01\ 0\ 0;\ -0.01\ 0\ 0;\ 0\ 0.01\ 0]$ m, "Phase delay" as $[0\ 0\ \pi/2\ 0]$ rad, and "Relative amplitude delay" as $[1\ 1\ 1\ 1]$. The predicted radiation force and torque are validated by the numerical method, as illustrated in Figs. 4 and 5. The interdistance is set to 20 mm to reduce the number of mesh elements in the full three-dimensional numerical model, which greatly saves simulation time. The particle is fixed in $[0\ 0\ 0]$ position, while its orientation varies along $x$-axis or $y$-axis, i.e., $\theta_{\text{rot},x} \in [0, 90°]$ or $\theta_{\text{rot},y} \in [0, 90°]$.

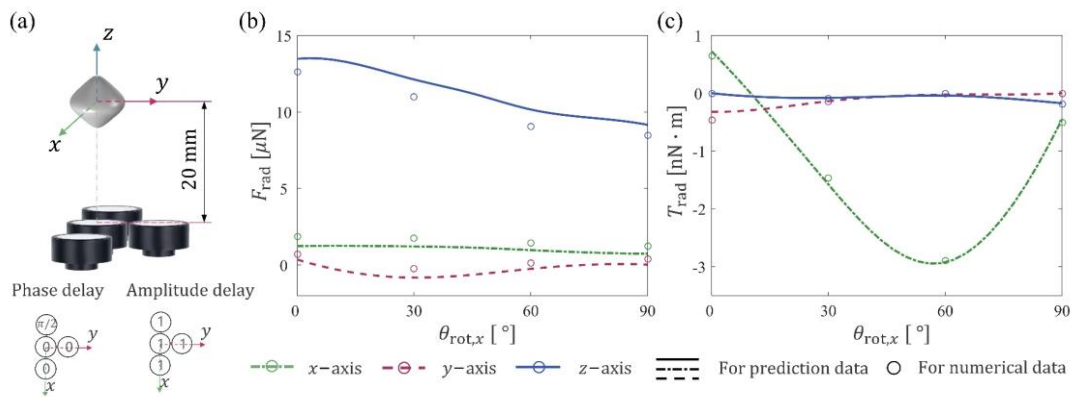

**Figure 4:** Theoretical and numerical calculations of the acoustic radiation force, $\vec{F}_{\text{rad}}$, and torque, $\vec{T}_{\text{rad}}$, acting on a sound-hard diamond particle (averaged radius of 2 mm) as a function of the particle orientation $\theta_{\text{rot},x} \in [0, 90°]$ in a four-transducer system. (a) Schematic diagram of the particle-transducer system and the phase and amplitude distributions of the transducer array. (b) and (c) Comparison plots for the radiation force and torque, respectively. The meanings of the circle marks and the curves are the same as those described in Fig. 2.

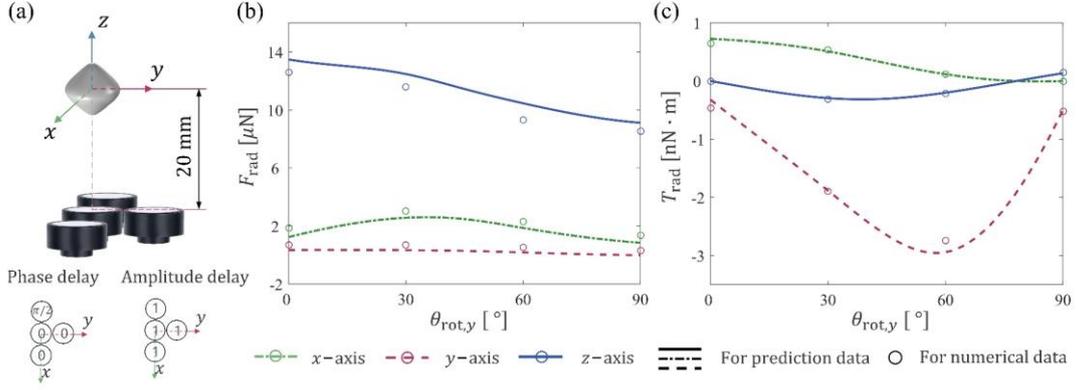

**Figure 5:** The same as in Fig. 4, but the particle orientation $\theta_{\text{rot}}$ is varied along $y$-axis.

It can be found in Figs. 2, 3, 4, and 5 that the acoustic radiation force and torque between our method and the numerical method are perfectly matched in various scenarios. However, it should be emphasized that for the cases using transducer array, since the distance between the scatterers and the transducer array is relatively small (i.e., the interdistance is 20 mm), the wavefield around the scatterers does not meet the far-field requirements, which compromises the prediction accuracy shown in Figs. 4 and 5. Furthermore, it is worth mentioning that Soundiation merely takes about ten seconds to estimate the radiation force and torque, while the numerical counterparts will cost about ten minutes (CPU: Intel i7-6700HQ 2.6 GHz; Maximum memory usage: 16 GB). Hence, the software becomes significant because it shows high computational accuracy and good computational robustness, while requiring much less computational time. Additionally, the validated COMSOL model, "Radiation_force_torque.mph", is accessible from the "Soundiation-Acoustophoresis ./docs/COMSOL" folder.

### 3.2. Prediction of the acoustophoretic process

In this section, we illustrate another functionality in the prediction of the acoustophoresis of an ellipsoidal particle ($c_n = [0.002\ 0\ 0.0004]$) above a transducer array using Soundiation. The ellipsoidal particle is rigid (sound-hard or Neumann boundary condition) with a density of $\rho_p = 15$ kg/m³, and its averaged radius is also set to $a = 2$ mm. In this way, the gravity can be calculated by $F_G \approx \frac{4}{3}\pi a^3 \rho_p g$, where $g$ is the acceleration of gravity. The ellipsoidal particle is shown in Fig. 6(a). The amplitude of the radial velocity of the transducers is also set to 1.5 m/s, with a vibration frequency of 40 kHz. The transducer array consists of five transducers with radii of 5 mm. The transducer configuration, phase delay, and relative amplitude delay can refer to Figs. 6(a) and 6(b). The initial particle position is given at [2 0 0] mm of Cartesian coordinates, while the particle rotates along $x$-axis for $\pi/6$, or $\theta_{\text{rot},x} = \pi/6$, as the initial particle orientation. The interdistance is set to 20 mm. The ending time and the time step are set to 0.1 s and 0.0001 s, respectively.

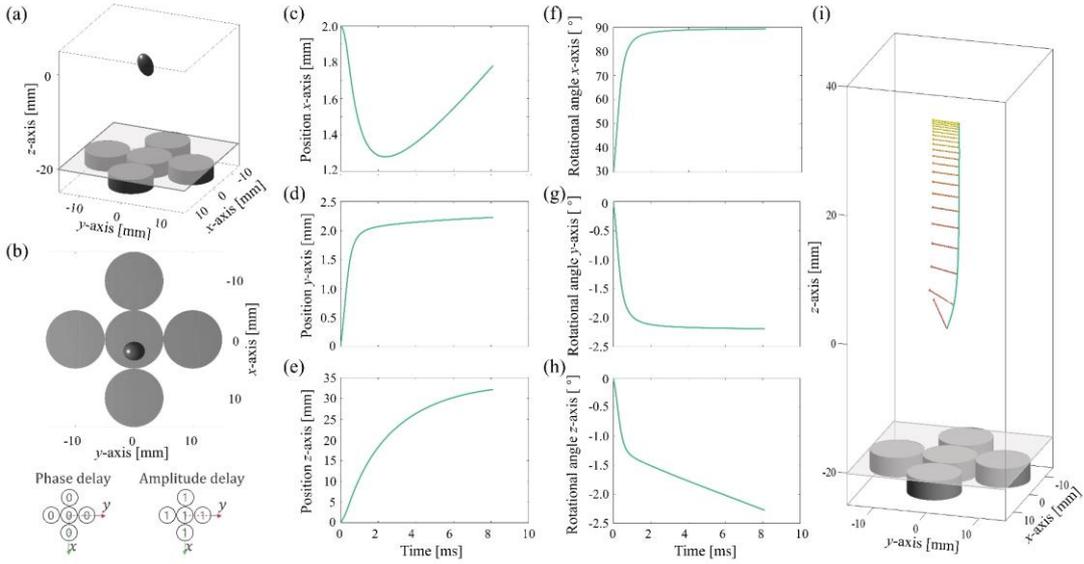

**Figure 6:** The time-variant translational and rotational trajectories of a sound-hard ellipsoidal particle with an averaged radius of 2 mm, manipulated by a transducer array. (a) The initial status of the particle-transducer system. The particle is placed at $(x, y, z) = [2, 0, 0]$ mm, rotating along $x$-axis with $\pi/6$. The interdistance is set to 20 mm. All transducers operate with the same phase and vibration amplitude. (b) The top view of the particle-transducer system (a). (c), (d), and (e) The translational trajectories of the particle along $x$-, $y$-, and $z$-axes, respectively. (f), (g), and (h) The rotational trajectories of the particle along $x$-, $y$-, and $z$-axes, respectively. (i) The time-variant translational trajectory (solid line) and particle orientation (arrow) visualized in three-dimensional. The color of these arrows represents the axisymmetric axis of the particle, while the color of the arrow is used to represent the increase of time (red-yellow color spectrum). The time intervals represented by any adjacent arrows are the same. These are 20 arrows showing the position and orientation of the particles from the start of the calculation (red arrow, 0 s) to the end of the calculation (yellow arrow, 0.008 s).

Figure 6 shows that the time-variant translational and rotational trajectories of an ellipsoidal particle can predict using the software, indicating that it is able to apply in describing the motion of various objects with a step-by-step and user-friendly guide approach. The prediction ends at 0.008 s (before 0.1 s) since the error of the positions and the rotation angles among two adjacent time steps are less than 5%.

# 4. Impact

Soundiation is a GUI-based software that allows users to design arbitrary axisymmetric geometries and evaluate the radiation force and torque exerted on them. Furthermore, the software supports user-customized incident wavefronts, including the transducer configurations and corresponding transducer parameters. Based on the derived radiation force and torque, taking into account the effects of viscosity, the dynamics of a non-spherical particle can be predicted and visualized. The above functionalities are beneficial for researchers who study axisymmetric micro- or nano-objects manipulation using the radiation force and torque in bioengineering [22][23] and medical [24][25] industries.

In addition to being suitable for a wider range of working parameters and application scenarios, Soundiation exhibits superior computational accuracy, high geometric adaptivity, and good robustness to various geometric features, while the computational efficiency is much higher than that of the full numerical methods. The time-consuming numerical simulations make the time-variant acoustophoretic process unpredictable, while is possible in Soundiation since the computational time typically takes about ten seconds per time step in common personal computers. Moreover, the user-friendly GUI-based application facilitates effortless use of the software by researchers unfamiliar with the theoretical principles.

Furthermore, the open-source codes allow researchers to add their features under the current framework and test new algorithms for other projects. For example, while the software merely supports the plane wavefronts or the wavefronts of circular radiators (or transducer array) in the current version, it can be simply extended to user-dependent arbitrary wavefronts through defining the proper wave functions in the source code "wave_function.m".

# 5. Conclusions

Soundiation is software that computes the radiation force and torque on user-specified axisymmetric particles under a user-customized incident wavefield (plane-wave mode or transducer-array mode). Geometric data for user-designed particles is automatically saved and can be imported directly into commercial numerical software (such as COMSOL Multiphysics) when needed to validate the predictions. It can be found that the predicted data are in perfect agreement with the direct numerical simulations, while the computational efficiency is much higher. Driven by radiation force and torque, the dynamics (or acoustophoresis) of any user-defined particle above the user-customized transducer array can be predicted and visualized. Additionally, the user-friendly GUI grants the users many advantages, making the software easier to use and extend, more efficient, and significantly reducing the learning curve for new users. The proposed software can be an effective tool to predict the motions of various non-spherical objects, helping to understand the acoustophoretic phenomena with a wide range of parameters in the biochemical industry.

# Current code version

| Nr. | Code metadata description | Please fill in this column |
|---|---|---|
| C1 | Current code version | v1.0 |
| C2 | Permanent link to code/repository used for this code version | https://github.com/Tountain/Soundiation-Acoustophoresis |
| C3 | Permanent link to Reproducible Capsule | |
| C4 | Legal Code License | GNU GENERAL PUBLIC LICENSE V3.0 |
| C5 | Code versioning system used | GitHub |
| C6 | Software code languages, tools, and services used | MATLAB |
| C7 | Compilation requirements, operating environments & dependencies | MATLAB 2010a and above |
| C8 | If available Link to developer documentation/manual | https://github.com/Tountain/Soundiation-Acoustophoresis/tree/main/docs/user_manual |
| C9 | Support email for questions | tianquan@connect.hku.hk |